# Strain Relaxation in Core-Shell Pt-Co Catalyst Nanoparticles

E. Padgett[a,*], M. E. Holtz[a,c], A. Kongkanand[b], D. A. Muller[a]

[a] School of Applied and Engineering Physics, Cornell University, Ithaca, New York 14850, USA
[b] Fuel Cell Business, Global Propulsion Systems, General Motors, Pontiac, Michigan 48340, USA
[c] Department of Metallurgical and Materials Engineering, Colorado School of Mines, Golden, Colorado 80401, USA
[*] Current address: National Renewable Energy Laboratory, Golden, Colorado 80401, USA

**Abstract**

Surface strain plays a key role in enhancing the activity of Pt-alloy nanoparticle oxygen reduction catalysts. However, the details of strain effects in real fuel cell catalysts are not well-understood, in part due to a lack of strain characterization techniques that are suitable for complex supported nanoparticle catalysts. This work investigates these effects using strain mapping with nanobeam electron diffraction and a continuum elastic model of strain in simple core-shell particles. We find that surface strain is relaxed both by lattice defects at the core-shell interface and by relaxation across particle shells caused by Poisson expansion in the spherical geometry. The continuum elastic model finds that in the absence of lattice dislocations, geometric relaxation results in a surface strain that scales with the average composition of the particle, regardless of the shell thickness. We investigate the impact of these strain effects on catalytic activity for a series of Pt-Co catalysts treated to vary their shell thickness and core-shell lattice mismatch. For catalysts with the thinnest shells, the activity is consistent with an Arrhenius dependence on the surface strain expected for coherent strain in dislocation-free particles, while catalysts with thicker shells showed greater activity losses indicating strain relaxation caused by dislocations as well.

**Introduction**

Proton exchange membrane fuel cells (PEMFCs) are increasingly expected to play a key role in decarbonizing heavy duty transportation, where long range, fast refueling, and carbon-free fuel are important advantages.[1] In comparison to light duty applications such as personal vehicles, fuel cells for heavy duty commercial applications must minimize overall lifetime costs, making both durability and efficiency key priorities.[1,2] The cathode catalyst layer, which performs the kinetically slow oxygen reduction reaction (ORR), is a critical component for the performance, efficiency, cost, and durability of PEMFCs. ORR kinetics are the greatest source of voltage losses for well-optimized membrane electrode assemblies (MEAs),[3] limiting the overall cell efficiency and requiring the use of scarce Pt-based catalysts to provide high power density.

Furthermore, degradation of the cathode catalyst layer limits the lifetime of PEMFCs and leads to gradual loss of efficiency and power density.[4]

Pt-based nanoparticles alloyed with a secondary transition metal such as Co or Ni are the most promising catalysts demonstrated to be viable in MEAs for automotive applications.[5] Pt-alloy catalysts such as Pt-Co or Pt-Ni provide significant enhancements in specific ORR activity over comparable pure Pt particles or surfaces, improving fuel cell performance and efficiency.

In practice, Pt-alloy nanoparticles generally have a core-shell structure when operating, as the acidic fuel cell environment dissolves the secondary non-precious metal near the catalyst structure, leaving a Pt shell surrounding an alloy core. Typically the catalyst is prepared by dealloying[6–9] through acid leaching to form the core-shell structure prior to the assembly of the MEA. This limits in-situ dissolution of the secondary metal, which leads to cationic contamination of the membrane. During the lifetime of the fuel cell, redeposition of dissolved Pt and leaching of the secondary metal causes the thickness of the Pt shell to grow up to several nanometers.[10,11]

Two mechanisms are generally posited for the ORR activity enhancement in Pt-alloy catalysts – strain or electronic charge transfer ("ligand effects"). Each of these mechanisms would alter the binding strength of adsorbed intermediate species in the ORR by altering the electronic structure of the Pt surface, as described by the Hammer-Norskov d-band model.[12–15] However, the Pt shell for operating catalysts is generally thicker than a single monolayer, making electronic charge transfer unlikely to impact the surface electronic structure because of the short electronic screening length in metals.[10,a] Thus while charge transfer effects may be important for some ORR catalysts, strain is the more probable mechanism under normal circumstances for Pt alloy catalysts. The importance of strain in surface reactivity has been understood theoretically for some time and validated for ideal surfaces.[14,15] In the case of Pt-Co and Pt-Ni ORR catalysts, the alloy core has a smaller lattice constant than pure Pt, resulting in a compressive strain exerted on the Pt shell. In the d-band model, the compressive strain is expected to broaden the Pt d-band, pushing its center down away from the Fermi level and weakening the bonding strength of oxygen intermediates.

The expected enhancement of ORR activity is observed for Pt-Co and Pt-Ni catalysts, and the advantage in activity over pure Pt is lost as the Pt shell grows in thickness due to electrochemical aging. However, few details are understood about the connections between structure, strain, and activity for Pt-alloy fuel cell catalysts and how the intended strain-induced activity is lost from catalyst degradation over the lifetime of the fuel cell. Macroscopic averages of the strain estimated by X-ray diffraction show a correlation between strain and activity but are in poor qualitative agreement with theory.[16] In particular, the strain effects expected for the spherical core-shell geometry typical of catalyst nanoparticles are not widely understood, and the differences from more widely understood planar films not always appreciated. Furthermore, fuel cell catalysts operating in a MEA environment are complex, with heterogeneous structures and many effects contributing to the observed bulk activity that are challenging to disentangle. It is important to understand real-world strain effects and their role in ORR activity to enable the strategic development of durable, high performance catalyst materials.

---

[a] Thomas-Fermi screening length of ~0.6Å for Pt is expected, compared to 2.8Å inter-atomic spacing.

Progress on this front has been limited by lack of characterization tools suitable for measuring strain in realistic fuel cell catalyst nanoparticles. Nanoscale strain characterization has been demonstrated using techniques including atomic resolution imaging and nanobeam electron diffraction (NBED), but prior demonstrated methods are poorly suited to the complexity of supported core-shell nanoparticle catalysts. To be effective for this system, a strain characterization technique must be able to provide sub-nm spatial resolution and be robust to varying specimen thickness, background, and crystal orientation to provide high throughput for statistically representative measurements. As shown in a prior publication,[17] NBED with the exit wave power cepstrum (EWPC) transform is a powerful technique to overcome these challenges.

In this investigation, we apply the NBED-EWPC technique alongside a simple continuum elastic model for core-shell particles to explore the basic strain effects relevant for Pt-M nanoparticle catalysts. We examine the strain distributions present in dealloyed and electrochemically aged catalysts. Two mechanisms for strain relaxation are identified: geometrical relaxation from Poisson expansion in spherical shells and dislocation-driven relaxation. We demonstrate that consideration of these two mechanisms can account for the loss of catalytic activity observed in a controlled series of catalysts dealloyed to different shell thicknesses.

**Materials and Methods**

*Catalyst materials and MEA assembly*

The Pt-Co catalysts used in this study were synthesized using an impregnation method followed by chemical dealloying. The catalyst was loaded on the carbon black support at 30wt% Pt. Both Vulcan and HSC carbon supports were used, and these catalysts and supports are described in more detail in a prior publication.[11] The electrochemically aged particles are sampled from both Pt-Co/HSC and Pt-Co/Vulcan catalysts, while the series of dealloyed catalyst particles are supported on HSC.

Membrane electrode assemblies (MEAs) were fabricated with the electrocatalysts of interest on the cathode. MEAs had an active area of 50 $cm^2$ and Pt loadings of 0.1 and 0.025 $mg_{Pt}/cm^2$ for the cathode and anode, respectively. Perfluorosulfonic acid (PFSA) Nafion® D2020 was used in the electrode at an ionomer to carbon weight ratio of 0.8 and 18 μm thick PFSA-based membranes were used. A 240 μm thick carbon paper with a 30 μm thick microporous layer (MPL) coated on top was used as the gas diffusion layer (GDL). The MEAs were fabricated using a catalyst-coated-membrane approach following a lamination procedure discussed in detail elsewhere.[3]

The series of dealloyed catalysts, with properties summarized in Table 1, was prepared starting with a dealloyed $Pt_{3.2}Co_1$/HSC catalyst with a shell thickness of ~0.5 nm. Prior to the initial dealloying the alloy composition was approximately $Pt_1Co_1$. The $Pt_{3.2}Co_1$/HSC catalyst was subjected to additional dealloying via an acid leach in 1 M nitric acid at 90°C for 1 day and 3 days, resulting in compositions of $Pt_{3.8}Co_1$ and $Pt_{4.8}Co_1$, respectively. A portion of the $Pt_{4.8}Co_1$/HSC catalyst was annealed at 400C for 1h in 5%$H_2$ in $N_2$ and dealloyed via acid leaching again for 1 day, resulting in a composition of $Pt_{5.5}Co_1$. A Pt/HSC catalyst was also prepared and annealed to have a particle size approximately equal to that of the Pt-Co catalysts.

Table 1: Summary of properties of the series of dealloyed catalysts.

| Catalyst | ECSA ($m^2/g_{Pt}$) | Mass Activity ($A/mg_{Pt}$) | Specific Activity ($\mu A/cm^2_{Pt}$) |
|---|---|---|---|
| $Pt_{3.2}Co_1$/HSC | 60 | 0.49 | 822 |
| $Pt_{3.8}Co_1$/HSC | 59 | 0.44 | 748 |
| $Pt_{4.8}Co_1$/HSC | 58 | 0.34 | 585 |
| $Pt_{5.5}Co_1$/HSC | 61 | 0.42 | 692 |
| Pt/HSC | 55 | 0.26 | 469 |

*MEA testing*

The catalysts in assembled MEAs were electrochemically aged following an accelerated stability test (AST) recommended by the U.S. Department of Energy (DOE)[18] consisting of 30,000 trapezoidal voltage cycles between 0.6 and 0.95 V with a 2.5 s dwell time at each voltage and a 0.5 s ramp time. The AST was performed at 80°C, 100% relative humidity, and ambient pressure. ORR activity measurements were performed in MEA and are reported at 0.9 $V_{RHE}$ at 80°C, 100% relative humidity, and 1 bar of $O_2$.[3] The electrochemically active surface area (ECSA) of Pt was measured by CO stripping in an MEA.[19]

*TEM sample preparation*

Samples of the electrochemically aged catalysts for (scanning) transmission electron microscopy ((S)TEM) were prepared by cross-sectioning the MEAs with an ultramicrotome. Strips cut from the MEAs were embedded in EMbed 812 Resin (Electron Microscopy Sciences) and cured at 60°C overnight. Sections were cut using a Leica Ultracut UCT Ultramicrotome at 40 – 50 nm thickness. Sections were collected on square-mesh Cu TEM grids with lacy carbon. Samples were cleaned with oxygen-argon plasma prior to TEM measurements.

*NBED measurements*

Nanobeam electron diffraction (NBED) mapping was performed in a FEI Titan Themis (S)TEM operated at 300 kV in STEM microprobe mode. Diffraction patterns were acquired on the EMPAD[20] pixelated detector, which has high dynamic range and single electron sensitivity to allow collection of all scattered electrons as well as the unsaturated direct beam. Typical acquisitions used a 1 ms exposure time with a beam current around 10-15 pA and 256 x 256 real-space pixels. Additional description of these methods can be found in a prior publication.[17]

*Calculation of strain maps*

All computation was performed in MATLAB. Lattice strain was calculated by tracking peaks in the exit wave power cepstrum (EWPC) calculated from the diffraction pattern using methods described in detail in a prior publication.[17] Strain maps were calculated for individual particles selected out of larger NBED maps, which typically include several particles. After peak maxima are located for the map, the 2 x 2 Lagrange strain tensor is calculated using the particle center as a reference. The strain ellipse axes are calculated from the eigenvalues of the strain tensor.

*STEM EELS Composition Maps and Shell Thickness Measurement*

STEM electron energy loss spectroscopy (EELS) measurements were performed on an FEI Titan Themis S/TEM operated at 300kV, equipped with a Gatan GIF Quantum spectrometer in single-range EELS mode. A convergence semi-angle of 21.4 mrad and a beam current of around 250 pA were used to acquire spectroscopic images with a pixel size of approximately 1 Å. Composition maps were extracted by integrating the signal from the Co $L_{2,3}$ and Pt $M_{4,5}$ edges after background subtraction using exponential and linear combination of power laws background fits, respectively. After integration, EELS maps were drift-corrected using an affine correction determined by comparing the simultaneous ADF signal to the fast-scanned overview image. The affine correction was constrained to only shear and stretch components consistent with uniform drift during STEM imaging. To measure the distribution of shell thicknesses for PtCo particles, the boundaries of the PtCo core and overall particle were segmented in the Co and Pt EELS maps, respectively, using an active contour method. The Pt shell thickness was calculated at each pixel on the particle surface as the distance from the nearest pixel in the PtCo core. Calculations were performed in Matlab using functions in the image processing toolbox. This method was described in more detail and validated in a previous report.[11]

**Elastic Continuum Theory for Core-Shell Particles**

Elastic continuum theory provides an analytically tractable approach to describe the surface strain expected for core-shell particles under simple assumptions. Here we investigate an analytical model for an ideal spherical core-shell particle with isotropic elastic properties and a coherently strained shell. This model provides basic intuition for understanding strain effects and is a reasonable set of assumptions for some simple catalysts. However, it ignores atomistic effects, which are likely to be salient for Pt-Co catalysts with structures on the order of a few atomic spacings, and does not account for crystal defects such as grain boundaries and dislocations. Variations in the particle geometry, including non-spherical particles or uneven shell thickness, will also cause the strain to deviate from this simple model, although these effects could be investigated numerically within the assumptions of elastic continuum theory by finite element modelling.

A detailed derivation of the core-shell model, starting from published calculations for similar systems,[21–23] is presented in the Appendix, and here we will present the general setup and key results. We will consider a core-shell particle with total radius $R$ and core radius $r_c$. Our task is to determine the distribution of strain throughout the particle, and especially at the particle surface given the particle geometry, lattice mismatch between the core and shell, and the elastic moduli of the materials. For simplicity, this discussion refers to the Pt-M system with negative lattice mismatch and compressive strain. We also assume that the elastic moduli of the core and shell materials are the same.

The particle consists of two regions: the spherical core and the spherical outer shell. If the core and shell are considered in their relaxed state, a gap is present between the outer surface of the core and the inner surface of the shell because of the difference in lattice parameter. The condition of coherent strain is introduced by a boundary condition requiring that these surfaces are brought into contact. This creates an isotropic pressure in the core (Equation A.14) and a compressive strain on the shell. As a result of the spherical geometry, Poisson expansion in the shell causes the compressive strain to decay going outward across the shell, decreasing as the inverse cube of the distance from the particle center $r$ (Equations A.15-16). The strain profile depends only on the Poisson ratio $\nu$, the core-shell lattice mismatch $\delta = (a_c - a_s)/a_s$ for core lattice parameter $a_c$ and shell lattice parameter $a_s$, and the ratio of the core radius to the total radius $\chi = r_c/R$. $\chi^3$ is also the core volume as a fraction of the total particle volume, and $(1 - \chi^3)$ is the shell volume as fraction of the total particle volume.

We will refer to two useful conventions for the strain. The material strain, defined as $\varepsilon(\mathbf{r}) = a(\mathbf{r})/a_0(\mathbf{r})$, has the lattice parameter $a(\mathbf{r})$ referenced to the local relaxed lattice parameter $a_0(\mathbf{r})$. An alternative convention is the Lagrange strain, defined as $\varepsilon_L(\mathbf{r}) = a(\mathbf{r})/a(\mathbf{r_0})$, which is referenced to the strained lattice parameter at some reference point $\mathbf{r_0}$. While the material strain is conceptually straightforward and the relevant quantity for considering impacts on chemical bonding and catalytic activity, the Lagrange strain is a more experimentally accessible quantity because it does not require prior knowledge of the local relaxed lattice.

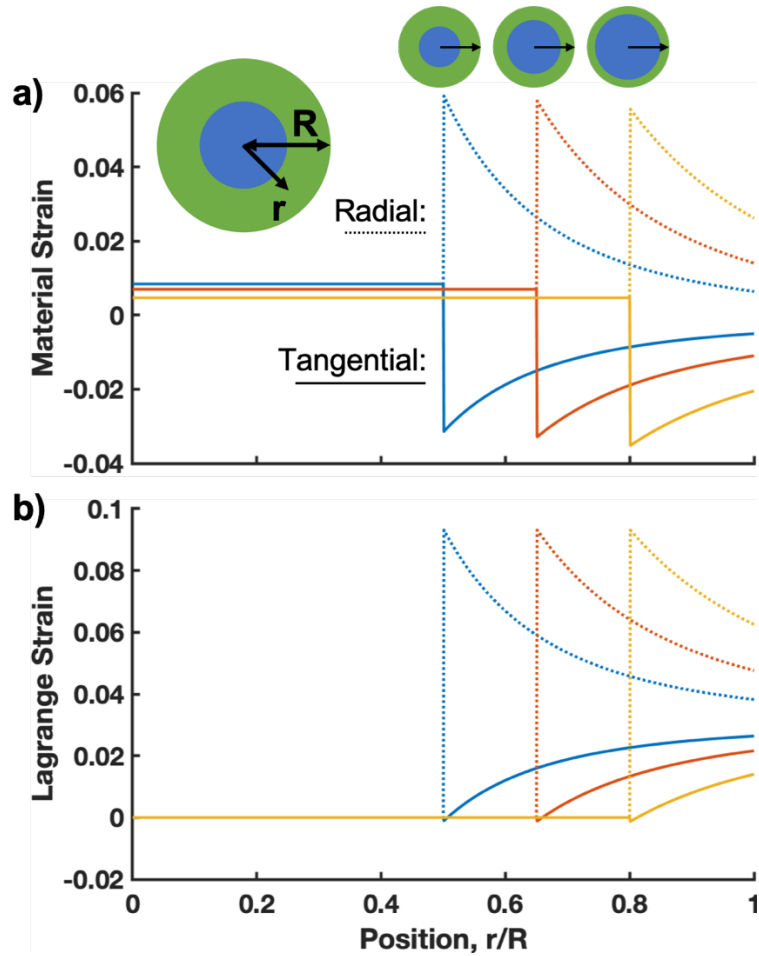

Figure 1: Theoretical tangential (solid) and radial (dotted) strain profiles for core shell particles with Poisson ratio $\nu=0.39$, lattice mismatch $\delta=-4\%$, and core-radius fractions $\chi=0.5$ (blue), 0.65 (red), 0.8 (yellow). The material strain (a) is referenced to the local relaxed lattice parameter, while the Lagrange strain (b) is referenced to the lattice parameter at the particle core. The core, at left, experiences a small expansive strain that decreases at higher $\chi$. The particle shell experiences a contraction in the tangential direction and expansion in the radial direction that decays with distance into the shell.

Figure 1 illustrates strain profiles for different values of $\chi$, and with $\nu=0.39$ and $\delta=-4\%$, corresponding approximately to a particle with a $Pt_1Co_1$ core and a Pt shell. The material strain profiles in (a) show that the core experiences an isotropic expansive strain exerted by the shell, which increases for thicker shells. The particle shell experiences a tangential compression and a corresponding radial Poisson expansion with an overall larger magnitude. Both tangential and radial strains relax gradually across the shell. Figure 1(b) shows the corresponding Lagrange strain profiles which are useful for interpretation of experimentally measured strain. The radial

Lagrange strain shows a prominent peak at the core/shell interface with a large magnitude, while the tangential Lagrange strain increases slowly in the shell and maintains a relatively small magnitude.

It is interesting to note that nowhere in the shell does the compressive tangential material strain equal the lattice mismatch. The strain magnitude is greatest at the core/shell interface, but there it is less than the lattice mismatch because of the expansion of the core exerted by the shell. Additional relaxation across the shell results in a tangential strain at the surface that is significantly lower than the lattice mismatch, especially for thicker shells.

For catalytically active nanoparticles, this model can quantify the material strain at the particle surface:

$$\varepsilon_s^{\theta\theta}(r=R) = \delta\chi^3, \tag{1}$$

$$\varepsilon_s^{rr}(r=R) = \frac{-2\nu}{1-\nu}\delta\chi^3. \tag{2}$$

The tangential strain $\varepsilon_s^{\theta\theta}$ is the product of the lattice mismatch and the core volume fraction $\chi^3$, while the radial strain $\varepsilon_s^{rr}$ also includes a Poisson-ratio correction with a numerical value of around 1.3 for Pt. This simple result is intuitively appealing. For the ideal case of coherent strain and uniform shell thickness the tangential strain is the same as the lattice mismatch averaged over the particle volume and can be determined from the average composition alone. The surface strain will only reach the core lattice mismatch in the limit of an infinitely thin shell, and as noted above, with the $\chi^3$ scaling decreases rapidly with increasing shell thickness.

**Experimental Results and Discussion**

*Observation of Strain Profiles in real Pt-Co Nanoparticle Catalysts*

In this study we have measured strain in practical Pt-Co fuel cell catalyst nanoparticles using nanobeam electron diffraction (NBED) mapping and the exit wave power cepstrum (EWPC) transform to allow robust calculation of local lattice distortions from the NBED patterns. This approach is discussed in detail in a previous publication.[17] Here we will begin by discussing the visualization and interpretation of strain distributions in Pt-Co nanoparticles and qualitative observations of the predictions of the elastic continuum model.

All experimentally measured strain presented here is the Lagrange strain, as there is no way to determine an ideal relaxed local lattice parameter. The reference point for the Lagrange strain is typically taken at the center of the particle, or at the center of the apparent core, if it differs from the particle center. Observed strain will be complicated by the fact that our measurements are a projection through particle and the finite size of the electron beam. Observed strain profiles should not be compared directly to the theoretical profiles shown in Figure 1, which are sampled at a 1D line in 3D space. Correcting for the projection will be considered in more detail in the following section, where we will interpret observed strain profiles by comparison with simulated profiles combining the elastic continuum model and a simple imaging model.

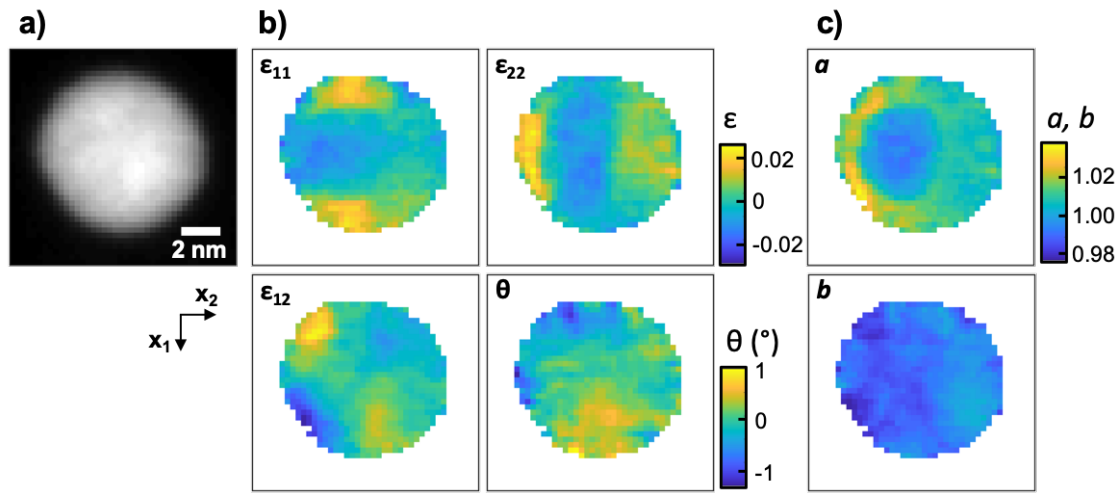

Figure 2: Projected strain maps for an electrochemically aged Pt-Co fuel cell catalyst nanoparticle. (a) Virtual annular dark field image of the particle calculated from the NBED dataset. (b) Maps of the lattice rotation $\theta$ and the strain tensor elements $\varepsilon_{11}, \varepsilon_{22}, \varepsilon_{12}$ in rectangular coordinates for the Lagrange strain referenced to the particle center. (c) Maps of the strain ellipse semi-major axis $a$ and semi-minor axis $b$, clearly showing the particle's core-shell structure.

Figure 2 shows projected strain maps for a single Pt-Co nanoparticle in the cathode of an electrochemically aged fuel cell MEA. As shown in the virtual annular dark field (ADF) image (a) the particle is roughly spherical with a ~8 nm diameter. Figure 2(b) shows maps of the Lagrange strain tensor elements $\varepsilon_{11}, \varepsilon_{22}, \varepsilon_{12}$ for vertical, horizontal, and diagonal strain, respectively, and the lattice rotation $\theta$. The $\varepsilon_{11}, \varepsilon_{22}, \varepsilon_{12}$ show that the particle is expanded radially outward near its edges, with high values of $\varepsilon_{11}$ at the top and bottom, high values of $\varepsilon_{22}$ at the left and right, and high or low values for $\varepsilon_{12}$ on the primary and secondary diagonals, respectively.

Because the direction of the strain changes around the shell, it is difficult to discern the underlying core shell structure of the particle from strain tensor components in rectangular coordinates. Expressing the strain in polar coordinates is also problematic because it requires identification of a center point, and depending on the geometry of each individual particle the choice of an appropriate center point may not be clear. An alternative formulation that avoids these challenges is the strain ellipse, which described in more detail in our prior publication.[17] The strain ellipse is formed conceptually by transforming a unit circle with the locally measured strain tensor. The semi-major axis $a$ and semi-minor axis $b$ of the strain ellipse are the distortion magnitude in the direction of most strain and the orthogonal direction of lowest strain, respectively. For compressively strained core-shell particles *a* is approximately the radial strain and *b* is approximately the tangential strain.[17]

Figure 2(c) shows maps of the strain ellipse axes for the example particle. The core-shell structure of the particle is clearly apparent in the semi-major axis map *a*. As in Figure 1(b), the core appears as a relatively flat, low strain region with a sharp high-strain ring around it at the core-shell interface. The core in this particular particle is offset to the left, leaving a relatively thick shell with more relaxed strain at the right and thin shell with high strain at the left.

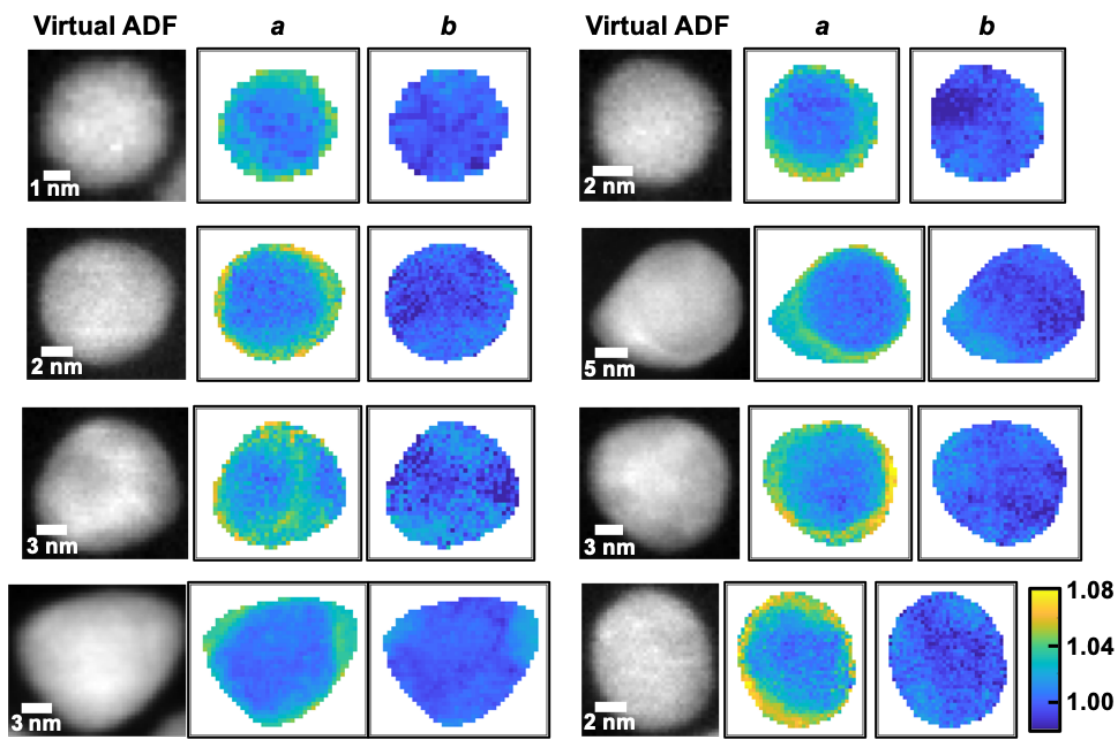

Figure 3: Strain maps for a selection of electrochemically aged Pt-Co fuel cell catalyst nanoparticles, including virtual annular dark field images, and maps of the strain ellipse semi-major axis *a* and semi-minor axis *b*.

One advantage of the NBED-EWPC strain mapping technique is the potential for relatively high throughput characterization, which is essential for allowing statistically representative study in intrinsically heterogeneous fuel cell catalyst materials. To illustrate this, Figure 3 shows a selection of 8 strain maps for Pt-Co particles in an electrochemically aged fuel cell cathode, including a virtual ADF image and the strain ellipse axis maps for each. The typical acquisition time per particle was in the range of 2-10 seconds. A large variety of strain distributions is observed in this selection, reflecting the variety of particle sizes, shapes, and internal structures. We observe relatively thin shells with high strain, and regions of thicker shells with significant strain relaxation. Thin shells on larger particles show higher strain than equally thin shells on smaller particles, reflecting the dependence of strain on the core radius fraction.

*Coherent Strain Relaxation*

To test the continuum elastic model's prediction of geometrically driven strain relaxation across shell we will make a quantitative examination of experimental strain profiles in shells of different thicknesses. As mentioned previously, the strain measured experimentally by NBED is not directly comparable to the predicted profiles shown in Figure 1 because of the effects of projection through the particle and blur due to the finite beam size. We account for these effects by the addition of a simple image model that integrates the strain parallel to the image plane across the thickness of the particle and smooths the profile with a Gaussian function matching the diffraction-limited beam size.

Figure 4 shows a comparison of experimental strain profiles for two particles to profiles simulated using the continuum elastic model and simple imaging model. Figure 4(a) shows a particle with a relatively thin Pt shell, with a strain profile sample along the magenta arrow plotted in (b). The tangential strain remains relatively low across the particle with a slight increase toward either end, while the radial strain increases significantly and peaks a short distance in from either surface. Figure 4(c) shows a simulated profile for a particle of the same size assuming a $Pt_1Co_1$ core and with a 1 nm shell thickness selected for consistency with the experimental profile in (b). The simulated profile including the imaging model is qualitatively consistent with and quantitatively very similar to the experimental profile. It is important to note that as a result of the imaging effects, the maximum observed radial strain is much lower than the maximum theoretical strain at the core/shell interface and is also displaced somewhat in position out into the shell. The observed profile at the core/shell interface is also considerably less sharp because a significant amount of strained shell is included in the projected measurement within the core.

The relaxation of strain across the Pt shell can be investigated by examining the strain in a region with a relatively thick shell, as shown in Figure 4(d), with the strain profile sampled at the magenta arrow plotted in (e). A simulated profile assuming a $Pt_1Co_1$ core and a 4 nm shell thickness is plotted in Figure 4(f). One caveat is that the shell thickness is not uniform across this particle, so the assumption of spherical symmetry made in the model is not obeyed and the profile is expected to differ somewhat from the model prediction. Thick but uniform shells were generally not observed in the particles examined in this study, with thick shells in limited areas being the norm, as shown in Figure 3. Nonetheless, the experimental and simulated profiles in Figure 4(e,f) are qualitatively quite similar, and the gradual relaxation of strain across the thick shell predicted by the model is observed in the experimental profile.

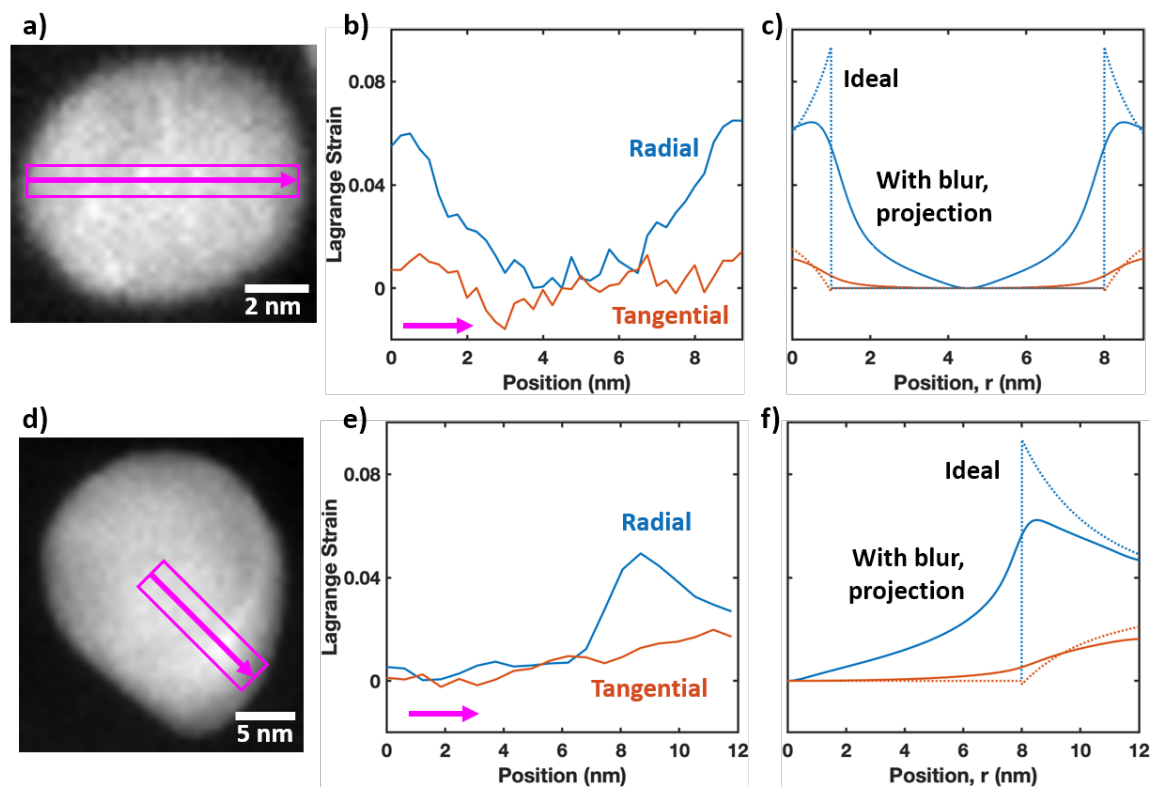

Figure 4: Line profiles of radial and tangential strain for electrochemically aged Pt-Co nanoparticles with a relatively thin Pt shell (a-c) and a relatively thick Pt shell (d-f). Virtual annular dark field images (a,d) of the nanoparticles with magenta arrow and box indicating the line profile regions used for calculation of the experimental strain profiles (b,e). Simulated profiles (c,f) were calculated for geometry similar to each experimental profile using the continuum mechanics model. The dotted lines show the strain along a single 1D line from the center to the edge of the 3D particle, while the solid lines include the effects of projection through the 3D particle and blurring from the finite beam size. Simulations assume a $Pt_{0.5}Co_{0.5}$ core and a pure Pt shell, with a thickness of 1 nm for (c) and 4 nm for (f).

These results demonstrate that despite the experimental complications in strain mapping Pt-Co nanoparticles using NBED, the maps can be interpreted to understand the underlying strain profiles by use of simple modeling. Furthermore, the consistency between the continuum elastic model and observed strain profiles, as well as the observed gradual relaxation of strain across thick Pt shells is evidence that the predicted geometric mechanism of strain relaxation is present in core-shell Pt-Co nanoparticles. This effect, which is not present in the more widely understood system of strained planar films, is important to consider for catalytic nanoparticles and other core-shell systems.

*Dislocation-driven Strain Relaxation*

In planar films, the primary mechanism of strain relaxation is the introduction of dislocations. This becomes energetically favored when the film thickness surpasses a critical thickness where the energy cost of the strain exerted on the film exceeds the energy cost of the added dislocations. Dislocations may play an important role in relaxing strain in core-shell nanoparticles as well, especially considering the large core-shell lattice mismatch sometimes present, for example ~4% in the case of our Pt-Co system. In particular, edge dislocations may be expected to appear at the core-shell interface with a missing plane of Pt atoms in the shell to relieve the compressive strain from the core.

Dislocations create distinctive strain fields[24] which can be used to recognize their presence in strain maps. Figure 5(a) illustrates the strain fields expected around an edge dislocation, which take a dumbbell-like shape. The strain in both the vertical $\varepsilon_{11}$ and horizontal directions $\varepsilon_{22}$ shows an expansion on the side of the missing atomic plane and a compression on the side of the extra plane. The shear strain $\varepsilon_{12}$ and lattice rotation $\theta$ each have dumbbells oriented perpendicular to those for $\varepsilon_{11}$ and $\varepsilon_{22}$ as the lattice bends around the dislocation core, although with opposite sign. However, as noted above, the strain tensor components in rectangular coordinates can be difficult to interpret in core-shell nanoparticles because of the changing direction of strain. Instead, the strain ellipse axes can show the regions of compression and expansion on either side of the dislocation, and the lattice rotation will show the expected dumbbell shape that should be easily recognized because the lattice rotation map would otherwise ideally be flat for a spherical core-shell particle.

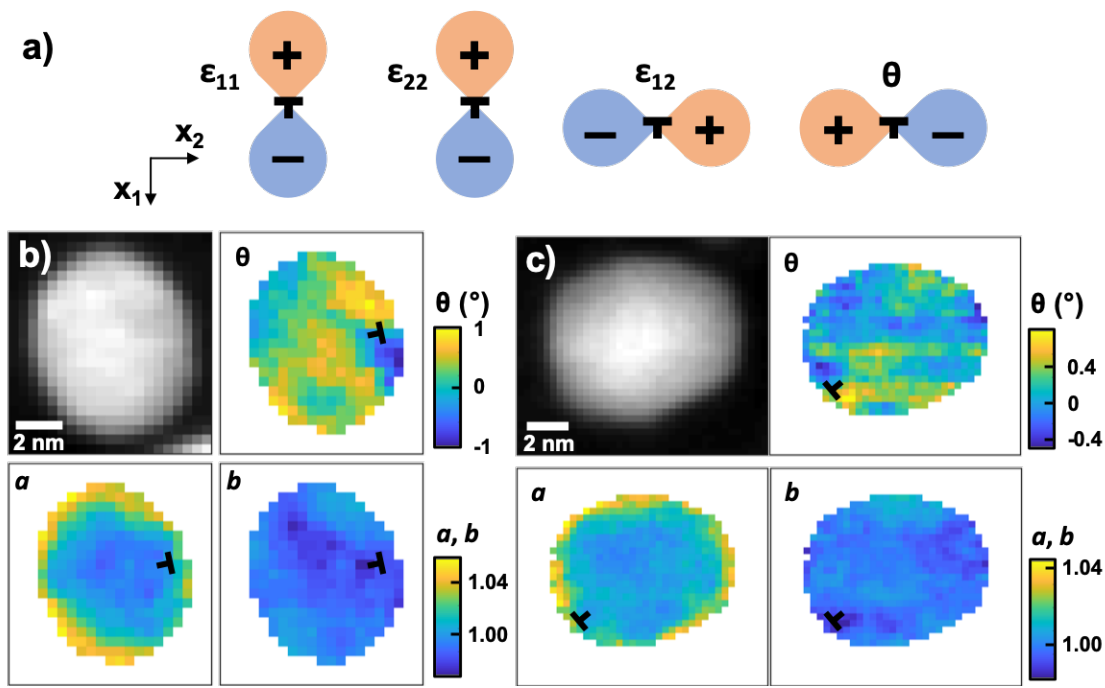

Figure 5: Observation of dislocations in core-shell Pt-Co nanoparticles. (a) Cartoon of the expected dipole-like form of the strain field tensor components $\varepsilon_{11}, \varepsilon_{22}, \varepsilon_{12}$ and lattice rotation $\theta$ around an edge dislocation, marked with the $\perp$ symbol. Strain maps of visible dislocations on an electrochemically aged particle (b) and a chemically dealloyed particle (c) including the virtual ADF image, lattice rotation $\theta$, and strain ellipse semi-major axis $a$ and semi-minor axis $b$.

Figure 5(b) and (c) show strain maps of Pt-Co particles with dislocations identified and marked with the $\perp$ symbol. The particle in (b) was electrochemically aged while the particle in (c) was subjected to an aggressive chemical dealloying to create a thicker Pt shell. In both cases, the dislocation appears to occur at the core-shell interface with a clear dumbbell visible in the lattice rotation $\theta$, a relaxation of strain in the nearby shell visible in the semi-major axis $a$, and a compressed point visible in the semi-minor axis $b$. Other locations on the particle show some of these signatures as well, although less clearly, suggesting that more dislocations may be present. Dislocations may have mixed screw and edge character, causing them to wind around the particle and potentially be less visible in strain maps from any particular angle. Quantifying the density or overall impact of dislocations is not a straightforward task because of their possibly complex 3D structure in core-shell nanoparticles. Nonetheless, these observations confirm that dislocations are present in Pt-Co nanoparticles whose shells have grown in thickness from either electrochemical aging or chemical dealloying and contribute to the relaxation of surface strain. Similar dislocations were not observed in the initial $Pt_{3.2}Co_1$/HSC sample.

*Rationalizing trends in ORR activity from strain microstructure*

The ultimate objective of this study is to relate the nanostructure and strain state of Pt-Co nanoparticles to their catalytic activity for oxygen reduction to expand the understanding that drives development of improved catalyst materials. Surface strain is a primary driver of catalytic activity, and the insights discussed in the previous sections can bring clarity to explain trends in activity that have previously been only vaguely understood.

The connection between strain and catalytic activity has been understood theoretically for some time from the d-band model.[12–15] By shifting the electronic d-band of the active metal, a change in surface strain is expected to give a linear change in the binding energy.[14,15] Under the assumption that the reaction has a single rate determining step (we are on the side of the volcano plot, and not the top) the catalytic activity, which is expected to follow an Arrhenius law depending exponentially on the energy barrier, will show an exponential Arrhenius dependence on the strain.

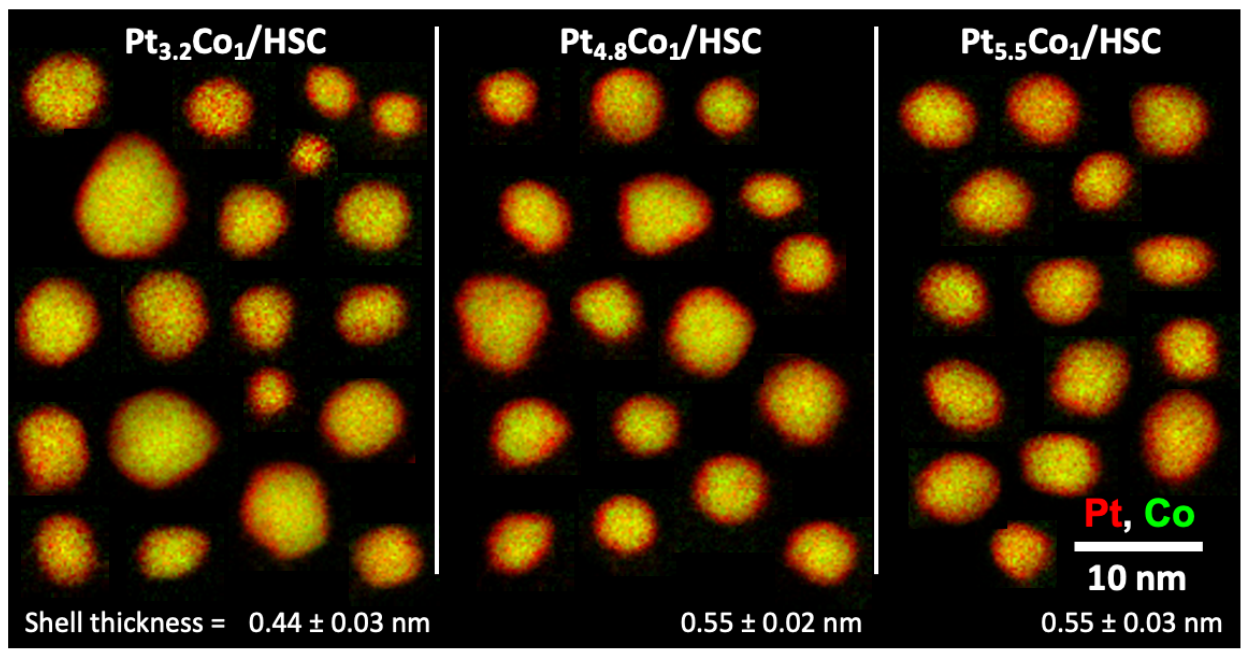

Figure 6: EELS maps used to characterize shell thickness for selected dealloyed Pt-Co nanoparticles. Platinum signal is shown in red, with cobalt signal in green, whith regions containing both appearing yellow. Measured average shell thickness are shown for each sample at the bottom of the figure, along with the standard error.

To examine the dependence of the activity on the microscopic strain state investigated in the previous sections, a series of Pt-Co/HSC catalysts was prepared with varying degrees of chemical dealloying intended to vary the Pt shell thickness. See Table 1 for a summary of the catalyst properties. Figure 6 shows EELS composition maps of three selected catalyst materials to illustrate the evolution of the core-shell structure, with quantitative measurements of the average shell thickness for each sample. The Pt-Co/HSC catalyst in particular was selected because of its highly uniform 4-5nm diameter size distribution, spherical morphology, and disinclination toward particle coalescence,[11] making it closely match the ideal model presented here. The initial catalyst has a composition of $Pt_{3.2}Co_1$ and was initially dealloyed to have a shell thickness of ~0.44 nm. Prior to the initial dealloying the alloy composition was approximately $Pt_1Co_1$. The $Pt_{3.2}Co_1$ catalyst was subjected to additional dealloying via acid leaching, resulting in compositions of $Pt_{3.8}Co_1$ and $Pt_{4.8}Co_1$. As shown in Figure 6, the further dealloying was measured to increase the Pt shell thickness by about 0.1 nm. This is somewhat less than the theoretically expected shell thickness change based on the composition (~0.2 nm), likely because the core-shell interface is somewhat diffuse, leading to some ambiguity in the EELS measurement. A portion of the $Pt_{4.8}Co_1$ catalyst was annealed to redistribute the cobalt and dealloyed again via acid leaching, resulting in a composition of $Pt_{5.5}Co_1$. Figure 6 shows that this resulted in a similar shell thickness to $Pt_{4.8}Co_1$, indicating instead that the composition change is due primarily to a decrease in the core Co content. A Pt/HSC catalyst was also prepared and annealed to have a particle size approximately equal to that of the Pt-Co catalysts. No significant difference in electrochemically active surface area (ECSA) was found for this collection of catalysts, implying that differences in activity due to particle size effects should be negligible. The specific oxygen reduction reaction (ORR) activity was measured for each of these catalysts and is compared to the expected strain in Figure 7.

The elastic continuum model presented above indicated that in the absence of dislocations (coherently strained interfaces only) the surface strain can be determined from the average composition alone (Equation 1) following a linear relationship. The theoretically expected surface strain, determined from the composition, of each of these catalysts is plotted in Figure 7(a). Note that the change in composition from $Pt_{3.2}Co_1$ to $Pt_{4.8}Co_1$, assuming an ideal geometry, is expected to produce only a small change in Pt shell thickness from ~0.5 nm to ~0.7 nm. This small change in shell thickness is expected to result in a significant ~40% strain relaxation from coherent geometric relaxation alone. However, the measured ORR specific activity decreases much more quickly after this dealloying than would be expected for an Arrhenius law dependence on the strain determined from the initial $Pt_{3.2}Co_1$ catalyst and the Pt reference. This outcome suggests that additional strain relaxation is taking place, which can likely be attributed to the introduction of dislocations into the thicker shells formed in the heavily dealloyed catalysts, as observed in Figure 5(c). Considering the high lattice mismatch of $\delta \approx -4\%$ between a $Pt_1Co_1$ core and Pt shell, dislocations are expected to be energetically favored above a critical thickness around only one unit cell. Furthermore, as shown in Figure 5(c), dislocations are experimentally observed in the shells of the chemically dealloyed particles, and it is therefore plausible to attribute the additional activity losses to dislocations in the Pt shell.

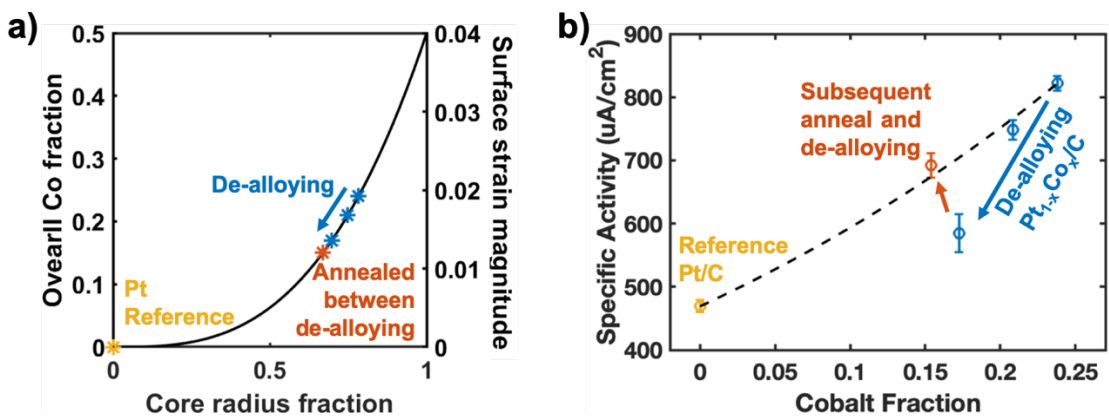

Figure 7: Correlation of theoretically expected strain to specific oxygen reduction reaction activity for a series of dealloyed Pt-Co catalysts. (a) Plot of theoretically expected relationship between core radius fraction $\chi$, average composition and tangential surface strain magnitude assuming coherent strain and a core composition of approximately $Pt_1Co_1$ and a -4% lattice mismatch. Stars mark points corresponding to the dealloyed Pt-Co series and a similarly sized Pt reference, with positions based on the average composition. (b) Plot of the experimentally measured specific oxygen reduction activity and average composition for the catalysts. The dotted line shows the expected trend assuming coherent surface strain that is linear with the average composition and activity following an Arrhenius law dependence on the strain between the Pt reference and the initial Pt-Co catalyst.

The activity recovers significantly, however, when the $Pt_{4.8}Co_1$ catalyst is annealed and dealloyed further by acid leaching reaching an overall composition of $Pt_{5.5}Co_1$. Notably, the ORR activity returns to the trend line expected for Arrhenius dependence with the expected coherent strain. The annealing process would be expected to uniformly redistribute the metal atoms in the catalyst particles and remove dislocations. The EELS measurements in Figure 6 indicate that the Pt shell thickness is similar in the $Pt_{4.8}Co_1$ and $Pt_{5.5}Co_1$ samples. This change toward a lattice mismatch from the lower Co content in the core makes the formation of dislocations less energetically favorable as well.

**Conclusion**

This investigation has explored basic strain effects in core-shell Pt-Co nanoparticle catalysts and their impact on ORR activity. An analytical model for the strain in core-shell particles derived from continuum elastic theory was presented and showed that for particles with coherently-strained, defect free core-shell interfaces, strain relaxes gradually across the particle shell as a result of Poisson expansion. The strain at the surface of coherently-strained particles is a simple function of the core-shell lattice mismatch and the relative size of the particle core, allowing the expected strain to be inferred from the average composition. Strain was characterized experimentally using NBED, which allows robust, high throughput strain measurements at sub-nm resolution. Experimentally measured strain profiles were generally consistent with profiles predicted from the elastic continuum model with projection and beam size effects taken into account, showing strain relaxation across thick Pt shells. Edge dislocations were also observed at the core-shell interface of particles that had been electrochemically aged or heavily acid leached. Strain relaxation in Pt-Co nanoparticle catalysts was thus shown to result from two mechanisms: geometric relaxation and dislocations. The continuum elastic model indicates that geometric relaxation produces the same loss of surface strain for particles of the same average composition regardless of their shell thickness, while dislocations are expected to be favored for thicker shells and a larger core-shell lattice mismatch. The impact of these two effects on ORR activity was investigated for a series of progressively dealloyed Pt-Co/HSC catalysts prepared to vary the shell thickness and core composition. For lightly dealloyed particles with thinner Pt shells, the activity followed a trend expected for Arrhenius dependence on the surface strain predicted for coherent strain with the continuum elastic model. However, for heavily dealloyed catalysts with relatively thick shells and a large lattice mismatch the activity fell faster than expected for coherent strain alone, indicating that dislocation-driven strain relaxation contributes to the activity loss as well for these catalysts.

**Acknowledgements**

This work was supported by the U.S. Department of Energy, Office of Energy Efficiency and Renewable Energy under grant DE-EE0007271. E Padgett acknowledges support from an NSF Graduate Research Fellowship (DGE-1650441). This work made use of electron microscopy facilities supported by the NSF MRSEC program (DMR 1120296) and an NSF MRI grant (DMR 1429155). The authors thank John Grazul, Malcolm Thomas, and Dr. Mariena Silvestry Ramos for assistance with electron microscopy facilities and sample preparation, and Paul Cueva for discussion of data analysis methods.

**Appendix: Derivation of Strain Distribution in Core-Shell Particles**

Here we describe a model of coherent strain in a uniform shell on an ideal, spherical core-shell particle using elastic continuum theory. Consider a core-shell particle with total radius $R$ and core radius $r_c$. Our task is to determine the distribution of strain throughout the particle, and especially at the particle surface – where it may impact chemical bonding and catalytic activity – given $R$, $r_c$, the lattice mismatch between the core and shell, the elastic moduli of the material. We will assume isotropic materials properties.

We will begin by following the analysis of a hollow spherical shell, subjected to internal and external pressures, described by Adel Saada in *Elasticity: Theory and Applications*.[21] We will present an abbreviated form of the mathematics, which Saada provides in more detail. This solution can be applied to the core-shell problem simply by determining the effective pressure to ensure contact between the core and the shell. This problem has been worked out in the literature[22,23,25] for core-shell quantum dots, where the core-shell structure exerts a pressure on the core that alters its band structure and the optical properties of the quantum dots. Our focus here is the impact of surface strain on the particle's catalytic activity and the microscopic profile of strain in the shell.

*Calculation of the displacement field*

Given its spherical geometry, the spherical shell problem is most easily approached in terms of the Lamé's strain potential $\varphi$ for the displacement field $\boldsymbol{u}$. The strain potential relates to the displacement field by $\boldsymbol{u} = \frac{1}{2G}\nabla\phi$, where $G$ is the shear modulus. With isotropic pressures, the system exhibits complete spherical symmetry, so only the radial displacement in the shell $u(r)$ must be considered (all other displacements are zero). The solution to the Navier equations has the form

$$\phi = \frac{C}{r} + Dr^2, \tag{A.3}$$

with constants $C$ and $D$, so the displacement is

$$2Gu(r) = -\frac{C}{r^2} + 2Dr. \tag{A.4}$$

The constants $C$ and $D$ can be solved for by setting the inner stress at $r = r_c$ equal to the inner pressure $P_i$ and the outer stress at $r = R$ equal to the outer pressure $P_o$. The displacement can then be determined as

$$u(r) = \frac{r(1+v)}{E}\left[P_i \frac{\left(\frac{1-2v}{1+v}\right) + \frac{1}{2}\left(\frac{R}{r}\right)^3}{\left(\frac{R}{r_c}\right)^3 - 1} + P_o \frac{\left(\frac{1-2v}{1+v}\right) + \frac{1}{2}\left(\frac{r_c}{r}\right)^3}{\left(\frac{r_c}{R}\right)^3 - 1}\right], \quad (A.5)$$

where $E$ is the Young's modulus and $v$ is the Poisson ratio. For a core-shell particle, $P_i$ is the contact pressure between the core and shell, while $P_o$ is the ambient pressure. Because the contact pressure is expected to be vastly larger than atmospheric pressure, for the purposes of strain calculation we may approximate $P_o \approx 0$, and consider only the contact pressure $P = P_i$. This simplifies the displacement in the shell to

$$u_s(r) = \frac{r(1+v)}{E}\left[P \frac{\left(\frac{1-2v}{1+v}\right) + \frac{1}{2}\left(\frac{R}{r}\right)^3}{\left(\frac{R}{r_c}\right)^3 - 1}\right]. \quad (A.6)$$

The definition of the pressure and displacement directions should be noted to ensure that the correct boundary condition is identified for core-shell contact. A positive pressure $P$ inside the spherical shell induces a positive displacement on the shell. The direction of the forces experienced by the core and the shell are opposite, so the core experiences a negative displacement for the same positive pressure:

$$u_c(r) = -\frac{P}{3K}r, \quad (A.7)$$

for bulk modulus $K$. Under this convention the pressure P is positive when the core is compressed while exerting a tensile stress on the shell. This will be the case for a positive lattice mismatch, defined as $\delta = (a_c - a_s)/a_s$ for relaxed lattice parameters $a_c$ and $a_s$ in the core and shell, respectively. Contact between the core and the shell enforces the boundary condition[25] on the core and shell displacements,

$$u_s(r_c) - u_c(r_c) = \delta r_c. \quad (A.8)$$

This condition allows us to solve for the contact pressure:

$$\delta r_c = \frac{r_c(1+v)}{E}\left[P \frac{\left(\frac{1-2v}{1+v}\right) + \frac{1}{2}\left(\frac{R}{r_c}\right)^3}{\left(\frac{R}{r_c}\right)^3 - 1}\right] + \frac{P}{3K}r_c$$

$$\frac{3K\delta}{P} = 1 + \frac{(1+v)}{(1-2v)}\left[\frac{\left(\frac{1-2v}{1+v}\right) + \frac{1}{2}\left(\frac{R}{r_c}\right)^3}{\left(\frac{R}{r_c}\right)^3 - 1}\right] = 1 + \left[\frac{\left(\frac{r_c}{R}\right)^3 + \frac{1}{2}\left(\frac{1+v}{1-2v}\right)}{1 - \left(\frac{r_c}{R}\right)^3}\right] = \left[\frac{\frac{3}{2}\left(\frac{1-v}{1-2v}\right)}{1 - \left(\frac{r_c}{R}\right)^3}\right]$$

$$P = 2K\delta \left(\frac{1-2v}{1-v}\right)\left(1 - \left(\frac{r_c}{R}\right)^3\right) = \frac{2K(1-2v)}{1-v}\delta(1-\chi^3). \tag{A.9}$$

For convenience, we have defined core radius fraction $\chi = r_c/R$. Note that $\chi^3$ is also the core volume as a fraction of the total particle volume, and $(1-\chi^3)$ is the shell volume as fraction of the total particle volume. The core pressure thus scales linearly with the lattice mismatch and with the shell volume fraction.

Substituting the pressure into the displacement relations yields

$$u_s(r) = \frac{2(1+v)}{3(1-v)}\delta\chi^3 r\left[\left(\frac{1-2v}{1+v}\right) + \frac{1}{2}\left(\frac{R}{r}\right)^3\right], \tag{A.10}$$

$$u_c(r) = -\frac{2}{3}\frac{(1-2v)}{1-v}\delta(1-\chi^3)r. \tag{A.11}$$

*Calculation of the strain fields*

The radial and tangential strains $\varepsilon^{rr}, \varepsilon^{\theta\theta} = \varepsilon^{\varphi\varphi}$ may be calculated from the displacement by the relations

$$\varepsilon^{rr} = \frac{du_r}{dr}, \tag{A.12}$$

$$\varepsilon^{\theta\theta} = \varepsilon^{\varphi\varphi} = \frac{u_r}{r}. \tag{A.13}$$

Here the strain $\varepsilon$ is the material strain, defined as $\varepsilon(r) = a(r)/a_0(r)$, with the lattice parameter $a(r)$ referenced to the local relaxed lattice parameter $a_0(r)$. An alternative convention is the Lagrange strain, defined as $\varepsilon_L(r) = a(r)/a(r_0)$, which is referenced to the strained lattice parameter at some reference point $r_0$. The Lagrange strain is experimentally straightforward to measure because it does not require prior knowledge of the local relaxed lattice.

In the particle core, there is a simple isotropic pressure-induced strain:

$$\varepsilon_c^{rr} = \varepsilon_c^{\theta\theta} = \varepsilon_c^{\varphi\varphi} = \frac{P}{3K} = -\frac{2(1-2v)}{3(1-v)}\delta(1-\chi^3). \tag{A.14}$$

In the particle shell, the strain varies spatially with radial position:

$$\varepsilon_s^{rr} = \left(\frac{(1+v)P}{E\left(\left(\frac{R}{r_c}\right)^3 - 1\right)}\right)\left[\left(\frac{1-2v}{1+v}\right) - \left(\frac{R}{r}\right)^3\right]$$

$$= \left(\frac{2(1+v)}{3(1-v)}\right)\delta\chi^3\left[\left(\frac{1-2v}{1+v}\right) - \left(\frac{R}{r}\right)^3\right], \tag{A.15}$$

$$\varepsilon_s^{\theta\theta} = \varepsilon_s^{\varphi\varphi} = \left(\frac{(1+\nu)P}{E\left(\left(\frac{R}{r_c}\right)^3 - 1\right)}\right)\left[\left(\frac{1-2\nu}{1+\nu}\right) + \frac{1}{2}\left(\frac{R}{r}\right)^3\right]$$

$$= \left(\frac{2(1+\nu)}{3(1-\nu)}\right)\delta\chi^3\left[\left(\frac{1-2\nu}{1+\nu}\right) + \frac{1}{2}\left(\frac{R}{r}\right)^3\right]. \tag{A.16}$$

Strain profiles for particles with different core radius fractions are illustrated in Figure 1.

*Strain at the particle surface*

The most interesting quantity for catalytically active nanoparticles is the strain at the particle surface:

$$\varepsilon_s^{rr}(r = R) = \frac{-3\nu}{E}\left(\frac{\chi^3}{1-\chi^3}\right)P = \frac{-2\nu}{1-\nu}\delta\chi^3 \tag{A.17}$$

$$\varepsilon_s^{\theta\theta}(r = R) = \frac{3(1-\nu)}{2E}\left(\frac{\chi^3}{1-\chi^3}\right)P = \delta\chi^3. \tag{A.18}$$

The tangential strain $\varepsilon_s^{\theta\theta}$ is the product of the lattice mismatch and the core volume fraction, while the radial strain $\varepsilon_s^{rr}$ also includes a Poisson-ratio correction. This simple result is intuitively appealing. For the ideal case of coherent strain and uniform shell thickness the tangential strain is the same as the lattice mismatch averaged over the particle volume and can be reasonably estimated from the average composition alone.